\documentclass[aps,prd,twocolumn,floatfix,nofootinbib,superscriptaddress]{revtex4-1}
\usepackage{dcolumn}

\usepackage{amsmath,amsfonts,extarrows,amssymb,mathrsfs,times,bm}
\usepackage{extarrows}
\usepackage{graphicx}
\usepackage{float}
\usepackage{graphicx,epstopdf}
\usepackage{dcolumn}
\usepackage{exscale}
\usepackage{relsize}
\usepackage{mathtools}
\usepackage{mhchem}
\usepackage[colorlinks=true,breaklinks=true,linkcolor=blue,citecolor=blue,urlcolor=blue]{hyperref}
\newcommand{\nn}{\nonumber}

\begin{document}

\title{The bending of a straight line}
\author{Zonghai Li}
\email[Electronic address:~]{lizzds@gznu.edu.cn}
\author{Xiao-Jun Gao}
\email[Electronic address:~]{gaoxiaojun@gznu.edu.cn}
\affiliation{School of Physics and Electronic Science, Guizhou Normal University, Guiyang, 550025, China}
\date{\today}
\begin{abstract}
In gravitational lensing under the weak-field approximation, the usual viewpoint is that light bending measures how a ray deviates from a straight line in Euclidean space. In this work, we take the opposite perspective: we ask how a straight line bends in a curved space, such as optical geometry—that is, how it deviates from geodesics. Using the Gauss–Bonnet theorem, we show that, at leading order, the deflection angle can be written as the integral of the geodesic curvature of a straight line in curved space. This reformulation emphasizes the global, coordinate-independent nature of the deflection angle and provides a complementary way of understanding the classical Gibbons–Werner method. To illustrate the idea, we apply it to three familiar spacetimes—Schwarzschild, Reissner--Nordstr\"om, and Kerr—and recover the well-known results. Furthermore, we extend the method to massive particles using the Jacobi metric, and illustrate it with the Reissner--Nordstr\"om spacetime.
\end{abstract}
\maketitle


\section{Introduction}

Geometry is not only a cornerstone of mathematics and a fundamental language of physics, but it is also indispensable to architecture, painting, and photography. It serves as a primary medium through which the deeper truths of existence are revealed to the human mind and remains a fundamental element of human culture.

In scientific exploration, one of the most remarkable examples of geometry shaping our view of the world is Einstein’s general relativity, which revealed the profound connection between spacetime and matter. Since Einstein, the geometrization of physics has advanced rapidly, with gauge field theories and string theory both deeply embedded in geometric frameworks. Beyond this well-established route, geometry offers a wealth of diverse and fascinating applications. For instance, in information geometry, geometric methods are used to explore the world of information~\cite{Amari-info-geo}; when extended to thermodynamics, this led to the field of thermodynamic geometry~\cite{Weinhold-1975,Ruppeiner-1995,Ingarden-1979,Janyszek-1990}, which has become an important tool in black hole thermodynamics~\cite{Ruppeiner2014,Wei-2019}. Differential geometric techniques also underpin the Kosambi–Cartan–Chern theory for analyzing the stability of dynamical systems~\cite{Antonelli-2003}. Furthermore, in quantum theory, geometry permeates diverse domains, including geometric phases~\cite{Berry-1984}, quantum computational geometry~\cite{Nielsen-2006-QIC,Dowling-2008-QIC} and its applications to holographic complexity~\cite{Jefferson-2017-JHEP,Susskind-2020}, geometric momentum on curved surfaces~\cite{Liu-2011-PRA}, and geometric methods in quantum control~\cite{Russell-2014,Brody-2015-PRL}.

Of course, in the geometric approach to physics, the geometrization of the principle of least action should not be overlooked, for it has a long history and continues to serve as a profound source of inspiration~\cite{Arnold-mech,Awrejcewicz-mech}. Fermat’s principle in optics states that light rays extremize the optical path length, offering one of the earliest variational principles with an explicit geometric interpretation. In mechanics, Maupertuis’ principle further demonstrates that the trajectories of conservative systems can be regarded as geodesics of Jacobi metric. These principles extend naturally to curved spacetime. Fermat’s principle provides a geometric formulation for describing light propagation in gravitational fields through the optical metric~\cite{Weyl,Perlick}. At the same time, the recent construction of a Jacobi metric in curved spacetime offers a new perspective on massive particle trajectories within the same variational–geometric setting~\cite{Gibbons2016,Chanda2019}. Using the purely spatial geometry of the optical or Jacobi metric, one can turn to differential geometry and the notion of intrinsic curvature to revisit the dynamics of light and massive particles, shedding fresh light on trajectories, stability, and the classical problem of black hole shadows~\cite{Das-EPJC-2017,Arganaraz-CQG-2021,Qiao-PRD-2022a,Qiao-PRD-2022b,Qiao-EPJC-2025,Cunha-CQG-2022,Bermudez-2025}. 

The effectiveness of this geometric perspective is most clearly manifested in gravitational lensing, a field of central importance for both testing fundamental theories and astronomical observations~\cite{Congdon-Keeton-2018}. A particularly elegant realization is provided by the Gibbons–Werner method, which applies the Gauss–Bonnet theorem within optical geometry to compute light deflection in the weak-field approximation, yielding a global and coordinate-independent result~\cite{Gibbons-Werner}. It should be noted that in stationary spacetimes, the optical metric takes the form of a Randers-type Finsler geometry, lying beyond the Riemannian framework. Werner~\cite{Werner2012} and, later, Ono–Ishihara–Asada~\cite{OIA2017} developed distinct approaches that allow the Gauss–Bonnet theorem to be applied in such situations. Through the Jacobi metric, the Gibbons–Werner method naturally extends from light rays to massive particles~\cite{massiveGB-CG,Crisnejo2019,massiveGB-CGJ,massiveGB-LiHZ,massiveGB-LiJa2020}, with notable applications to the lensing of charged particles in backgrounds involving both gravitational and electromagnetic fields~\cite{massiveGB-LiWJ,LiJia-2024}. In addition, alternative formulations of the Gibbons–Werner method have been introduced in order to treat diverse physical situations~\cite{ISOA2016,OIA-OA,TOA,massiveGB-LiZA,HuangCa,HuangCb,HuangCL,HuangSC}. Overall, this geometric approach has invigorated the field, giving rise to many interesting studies, such as~\cite{Jusufi-Ovgun-2018,Ono-Ishihara-Asada-2018,Ovgun-2018,Ovgun-Sakalli-Saavedra-2018,Ovgun-Sakalli-Saavedra-2019,Crisnejo-Gallo-Rogers-2019,Gao:2023ltr,Ali2025a,Ali2024b}. As a side note, this line of research has also stimulated fruitful interactions at the interface between mathematics and physics~\cite{Gibbons-PRD-2009a,Gibbons-PRD-2009b,Petters-Werner-GRG-2010,Werner-Obs-2010,Roesch-Werner-PAMQ-2020,Halla-thesis-2022}.

When viewed through the lens of the Gibbons–Werner method, the traditional geodesic approach—based on calculating the change of the coordinate angle—may be interpreted as computing the deviation of a light or particle trajectory from a straight line in Euclidean space. Equivalently, the total deflection angle is given by the integral of the geodesic curvature of the trajectory in flat space~\cite{Li-Zhou}. This raises a natural question: can we look at the problem from the opposite perspective? Instead of asking how light rays bend relative to straight lines in Euclidean space, can we ask how straight lines bend relative to geodesics in curved space? The purpose of this paper is to explore precisely this idea. To this end, we build on the geometric approach of Gibbons and Werner, applying the Gauss--Bonnet theorem within the framework of optical geometry and making use of straight lines to analyze light deflection.

The outline of this paper is as follows. Sec.~\ref{Czesław Miłosz} reviews the Gibbons--Werner method, providing the geometric background based on the Gauss--Bonnet theorem, which sets the stage for Sec.~\ref{Jorge Luis Borges}, where we present our approach that expresses the deflection angle through the geodesic curvature of a straight line in curved space. Sec.~\ref{Patrick Modiano} applies our formula to compute the light deflection in three representative spacetimes: Schwarzschild, Reissner--Nordstr\"om, and Kerr. Sec.~\ref{AIQING} extends the framework to massive particles, taking the Reissner--Nordstr\"om case as an explicit example.
Finally, Sec.~\ref{Marguerite Yourcenar} concludes the paper. Throughout, we adopt geometric units with $c=G=1$.

\section{A Review of the Gibbons-Werner Method}
\label{Czesław Miłosz}

To naturally introduce the present work, it is necessary to review the Gibbons--Werner method~\cite{Gibbons-Werner}. This section begins with the classical Gauss--Bonnet theorem for domains with boundary on surfaces, followed by a brief introduction to optical geometry. The Gauss--Bonnet theorem is then applied to optical geometry to derive the Gibbons--Werner formula for light deflection in the weak-field limit.

\subsection{The Gauss--Bonnet Theorem}

Let $D$ be a subset of a compact oriented two-dimensional Riemannian manifold, 
with piecewise smooth boundary $\partial D = \cup_i \partial D_i$. Denote by $K$ the Gaussian curvature of the surface and by $\chi(D)$ the Euler characteristic of $D$. The Gauss--Bonnet theorem can then be written as~\cite{doCarmo}
\begin{align}
	\label{GBT}
	\sum_i \psi_i + \sum_i \int_{\partial D_i} k_g \, d\ell
	+ \iint_D K \, dS = 2\pi \chi(D),
\end{align}
where $k_g$ denotes the geodesic curvature along the boundary, $\psi_i$ are the exterior angles at the corners with respect to the positive orientation, $d\ell$ is the line element along the boundary, and $dS$ is the area element of $D$. The three terms on the left-hand side correspond to curvature contributions of dimension 0, 1, and 2, respectively, while the right-hand side yields a topological invariant. Hence, the total curvature of $D$ is equal to a purely topological quantity, underscoring the deep interplay between geometry and topology.

\subsection{The Optical Metric}

Consider a static, spherically symmetric spacetime written in Schwarzschild coordinates 
$(t,r,\theta,\phi)$ as
\begin{align}
	\label{ssspacetime}
	ds^2 = -A(r)\, dt^2 + B(r)\, dr^2 + r^2 \left(d\theta^2+\sin^2\theta\, d\phi^2\right).
\end{align}
For null geodesics, setting $ds^2=0$ and solving for $dt^2$ yields
\begin{align}
	\label{layonha}
	dt^2 = \alpha_{ij}\, dx^i dx^j 
	= \frac{B}{A}\, dr^2 + \frac{r^2}{A}\left(d\theta^2+\sin^2\theta\, d\phi^2\right),
\end{align}
which defines a positive-definite Riemannian metric $\alpha_{ij}$, referred to as the 
\emph{optical metric}. Fermat's principle ensures that light rays correspond to geodesics 
in the optical geometry $(\mathcal{M}, \alpha_{ij})$.

Without loss of generality, we consider light propagation in the equatorial plane, 
in which case the optical space is denoted by $(\mathcal{M}^2,\alpha_{ij})$ with metric
\begin{align}
	dt^2 = \alpha_{ij}\, dx^i dx^j = \frac{B}{A}\, dr^2 + \frac{r^2}{A}\, d\phi^2.
\end{align}

\subsection{The Gibbons--Werner Formula}
\label{Gibbons-Werner-Formula}
\begin{figure}[!t]
	\centering
	\includegraphics[width=8cm]{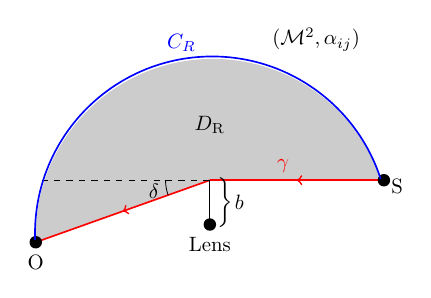}
	\caption{Region $D_R\subset(\mathcal{M}^{2},\alpha_{ij})$ bounded by the 
		photon trajectory $\gamma$ and the circular arc $C_R$. A light ray from the source $S$ 
		is deflected by the lens and received by the observer $O$, with impact parameter $b$ and 
		deflection angle $\delta$.}
	\label{XuWei_GW}
\end{figure}

Weak gravitational lensing is studied in the optical space $(\mathcal{M}^{2},\alpha_{ij})$, 
assumed asymptotically Euclidean. In this setting, an equatorial light ray $\gamma$ 
originates from the source $S$ in the asymptotic region, is deflected by the lens, and 
reaches the observer $O$, also in the asymptotic region, with a small deflection angle 
$\delta$, as shown in Fig.~\ref{XuWei_GW}.

Consider a nonsingular region $D_R \subset (\mathcal{M}^2, \alpha_{ij})$ with boundary 
$\partial D_R = \gamma \cup C_R$, where $C_R$ is the circular arc $r=R=\text{const}$.  
Applying the Gauss--Bonnet formula \eqref{GBT} to $D_R$ gives
\begin{align}
	\label{zai-Biechu}
	\psi_S + \psi_O + \int_{C_R}\kappa_g\, dt + \iint_{D_R}K\, dS = 2\pi\chi(D_R).
\end{align}
Since $\gamma$ is a geodesic, $k_g(\gamma)=0$, and for a nonsingular domain $D_R$ one has $\chi(D_R)=1$. 
In the limit $R\to\infty$, with $\phi_S=0$ and $\phi_O=\pi+\delta$, the exterior angles satisfy 
$\psi_S+\psi_O\to\pi$, and the boundary term reduces to $k_g(C_R)\,dt \to d\phi$. 
Consequently, Eq.~\eqref{zai-Biechu} becomes
\begin{align}
	\iint_{D_\infty} K\, dS + \int_0^{\pi+\delta} d\phi + \pi = 2\pi,
\end{align}
which gives the Gibbons--Werner formula~\cite{Gibbons-Werner}
\begin{align}
	\label{Gibbons-Werner}
	\delta = -\iint_{D_\infty} K\, dS.
\end{align}
This expression shows that the deflection angle is a global, coordinate-independent quantity.  

\subsection{Leading-order approximation}

The Gibbons--Werner formula \eqref{Gibbons-Werner} is an exact geometric expression for the weak-field deflection angle. In practice, however, its evaluation requires a small-parameter expansion. The standard approach is to treat the spacetime parameters (such as the mass \(M\), the spin \(a\), and the charge \(Q\), etc.) as small quantities, expand the metric, Gaussian curvature, and particle trajectory in powers of these small parameters, and compute the deflection angle order by order through an iterative procedure (see, e.g., Ref.~\cite{Werner2012}).

At leading order (i.e., the dominant order, referring to the lowest nonvanishing contribution in the expansion), it suffices to replace the actual light ray by the unperturbed straight-line trajectory \(r = b/\sin\phi\) with \(0 \leq \phi \leq \pi\), because the lowest-order nonvanishing contribution of the spacetime parameters is already encoded in the Gaussian curvature, while trajectory corrections contribute only at higher orders. Recall that the area element of the optical metric in polar coordinates is \(dS = \sqrt{\alpha}\, dr\, d\phi\), where \(\alpha = \det(\alpha_{ij})\) is the determinant of the optical metric. The Gibbons-Werner formula \eqref{Gibbons-Werner} can then be expanded to leading order as
\begin{align}
	\label{Michel Tournier}
	{}^{L}\delta 
	= - \int_{0}^{\pi}\!\int_{b/\sin\phi}^{\infty} 
	{}^{L}\!\big[K\sqrt{\alpha}\,\big]\, dr\, d\phi.
\end{align}
Here and throughout, the superscript \(L\) denotes the leading-order contribution.

\section{Bending of a straight line in curved space}
\label{Jorge Luis Borges}

\subsection{The deflection angle from the bending of a straight line}

\begin{figure}[!t]
	\centering
	\includegraphics[width=8cm]{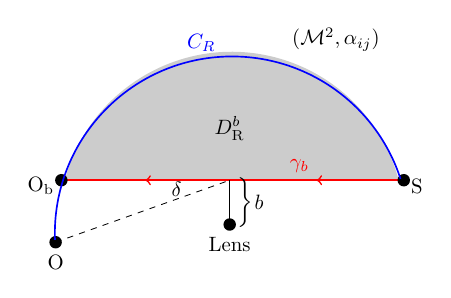}
	\caption{Region $D_R^b \subset (\mathcal{M}^{2},\alpha_{ij})$ bounded by the straight line 
		$\gamma_b$ and the circular arc $C_R$. The straight line from the source $S$ to the observer $O_b$ has nonvanishing geodesic curvature due to the lens.}
	\label{XuWei_sl}
\end{figure}

Here we consider the region $D_R^b \subset (\mathcal{M}^2,\alpha_{ij})$ with boundary 
$\partial D_R^b = \gamma_b \cup C_R$, where $C_R$ is defined as before. The curve $\gamma_b$ denotes straight line with trajectory $r = b/\sin\phi$, extending from the source $S$ to the point $O_b$ in the asymptotic region, as shown in Fig.~\ref{XuWei_sl}.

The region $D_R^b$ shares exactly the same topology and geometry as $D_R$ discussed in Sec.~\ref{Gibbons-Werner-Formula}. Note that $\gamma_b$ is not a geodesic of $(M^{2},\alpha_{ij})$, hence $k_g(\gamma_b)\neq 0$. Applying the Gauss--Bonnet theorem \eqref{GBT} to region $D_R^b$, we obtain:
\begin{align}
	\psi_S + \psi_{O_b} + \int_{C_R}k_g\, dt+\int_{\gamma_b}k_g\, dt + \iint_{D_R^b}K\, dS = 2\pi.\nn
\end{align}

As $R \to \infty$, setting $\phi_S=0$ and $\phi_{O_b}=\pi$, we have $\psi_S+\psi_{O_b}\to\pi$ and $k_g(C_R)\,dt \to d\phi$. In this limit, the above equation becomes
\begin{align}
	\pi+ \int_0^{\pi}d\phi+\int_{O_b}^{S}k_g(\gamma_b)dt+\iint_{D_{\infty}^{b}}  K d S=2\pi,\nn
\end{align}
which further implies
\begin{align}
	\int_{S}^{O_b}k_g(\gamma_b)dt=\iint_{D_{\infty}^b}  K d S=\int_{0}^{\pi} \int_{\frac{b}{\sin\phi}}^{\infty}K\sqrt{\alpha}~ drd\phi.\nn
\end{align}
Considering the leading-order approximation and using Eq.~\eqref{Michel Tournier}, we obtain
\begin{align}
	\int_{S}^{O_b} {}^{L}[k_g(\gamma_b)]\, dt 
	= \int_{0}^{\pi} \int_{\tfrac{b}{\sin\phi}}^{\infty} {}^{L}\!\left[K\sqrt{\alpha}\right] dr\, d\phi 
	= -{}^{L}\delta. \nn
\end{align}
Equivalently,
\begin{align}
	\label{Gary Snyder}
	{}^{L}\delta = -\int_{S}^{O_b} {}^{L}k_g(\gamma_b)\, dt .
\end{align}

	Eq.~\eqref{Gary Snyder} is the central result of the present work. It shows that, at leading order, the deflection angle can be expressed as the integral of the geodesic curvature of a fixed straight line in curved optical space. This expression makes explicit the global, coordinate-independent character of the deflection angle, offering a clear geometric interpretation. It provides a new perspective on gravitational deflection, complementing two existing approaches:
	
\begin{itemize}
		\item \textbf{The traditional geodesic method} solves the null geodesic equation to find the change in the coordinate angle \(\Delta\phi\). Geometrically, this is equivalent to describing the deviation of the bent light ray from a straight line in flat space~\cite{Li-Zhou}.
		\item \textbf{The classical Gibbons--Werner method} \eqref{Gibbons-Werner} expresses the deflection angle as the integral of the Gaussian curvature over the infinite region outside the light ray in optical geometry.
	\end{itemize}
	
	Thus, our method stands in a dual relation to the traditional geodesic method: whereas the latter describes the bending of light relative to a straight line in flat space, the former describes the bending of a straight line relative to geodesics in curved space. This perspective is new in the study of gravitational lensing. At the same time, it serves as a complement to the Gauss--Bonnet method: whereas the Gibbons--Werner formula expresses the deflection angle as the integral of Gaussian curvature over the infinite region outside the geodesic, our method expresses it as the integral of geodesic curvature along a fixed straight line.

Moreover, the deflection angle~\eqref{Gary Snyder} can be expressed as
\begin{align}
	\label{leadefangexp}
	{}^{L}\delta = -\int_{0}^{\pi} {}^{L}\!\left[k_g(\gamma_b)\left(\frac{dt}{d\phi}\right)\right] d\phi .
\end{align}

\subsection{Geodesic curvature of $r=\frac{b}{\sin\phi}$}

We consider a two-dimensional Riemannian optical metric  
\begin{align}
	\label{youziS}
	ds^2 = \alpha_{rr}(r)\,dr^2 + \alpha_{\phi\phi}(r)\,d\phi^2.
\end{align}

Along the curve $r=r(\phi)$, writing $r'=\frac{dr}{d\phi}$, the unit tangent vector is
\begin{align} \label{velocity} 
	\mathcal{T}^i =\left(\frac{dr}{dt},\frac{d\phi}{dt}\right) 
	= \frac{1}{\Xi}\,\left(r',\,1\right),
\end{align}
where $\Xi = \sqrt{\alpha_{rr}\,(r')^2 + \alpha_{\phi\phi}}$.

Along the curve, let $N^i$ be the unit vector in the tangent space orthogonal to $\mathcal{T}^i$, chosen so that $\{\mathcal{T}^i, N^i\}$ forms a positively oriented orthonormal frame. It then satisfies
\begin{align}
	\alpha_{ij}N^iN^j = 1, 
	\qquad 
	\alpha_{ij}N^i\mathcal{T}^j = 0,
\end{align}
and is explicitly given by
\begin{align}
	\label{norvec}
	N^i = \Bigl(-\sqrt{\tfrac{\alpha_{\phi\phi}}{\alpha_{rr}}}\,\mathcal{T}^\phi,\,
	\sqrt{\tfrac{\alpha_{rr}}{\alpha_{\phi\phi}}}\,\mathcal{T}^r\Bigr).
\end{align}

Then the geodesic curvature of the curve is defined as~\cite{Tu-diff-geo}  
\begin{align}
	\label{geodef}
	k_g = \alpha_{ij}\,\mathcal{A}^i N^j .
\end{align}
The acceleration vector takes the form  
\begin{align}
	\label{Marguerite Duras}
	\mathcal{A}^i = \nabla_{\mathcal{T}} \mathcal{T}^i
	= \frac{d\mathcal{T}^i}{dt} + \Gamma^i_{jk}\,\mathcal{T}^j \mathcal{T}^k ,
\end{align}
where $\Gamma^i_{jk}$ denote the Christoffel symbols associated with the metric $\alpha_{ij}$.

Substituting \eqref{norvec} into \eqref{geodef}, we obtain  
\begin{align}
	\label{geocur}
	k_g = \sqrt{\alpha}\,\bigl(\mathcal{A}^\phi \mathcal{T}^r - \mathcal{A}^r \mathcal{T}^\phi\bigr)
	= \sqrt{\alpha}\,\bigl(\mathcal{A}^\phi r'- \mathcal{A}^r \bigr)\frac{d\phi}{dt},
\end{align}
where $\alpha=\alpha_{rr}\alpha_{\phi\phi}$ is the determinant of the metric $\alpha_{ij}$, and Eq.~\eqref{velocity} has been used.

	Using the unit tangent components \eqref{velocity} and the relation \(d/dt = \Xi^{-1} d/d\phi\), we obtain
	\begin{align}
		\frac{d\mathcal{T}^r}{dt} = \frac{1}{\Xi^2} r'' - \frac{\Xi'}{\Xi^3}  r', \quad
		\frac{d\mathcal{T}^\phi}{dt} = -\frac{ \Xi'}{\Xi^3}, \label{dangddaiTT}
	\end{align}
	where \(\Xi' = d\Xi/d\phi\).
	
	For the diagonal optical metric \eqref{youziS}, the nonvanishing Christoffel symbols are
	\begin{subequations}
		\label{WuliXue}
		\begin{align}
			\Gamma^r_{rr} &= \frac{1}{2\alpha_{rr}}\frac{d\alpha_{rr}}{dr}, \\
			\Gamma^r_{\phi\phi} &= -\frac{1}{2\alpha_{rr}}\frac{d\alpha_{\phi\phi}}{dr}, \\
			\Gamma^\phi_{r\phi} &= \frac{1}{2\alpha_{\phi\phi}}\frac{d\alpha_{\phi\phi}}{dr}. 
		\end{align}
	\end{subequations}

	Substituting Eqs.~\eqref{dangddaiTT} and \eqref{WuliXue} into the acceleration \eqref{Marguerite Duras}, and evaluating all quantities on the curve \(r = r(\phi)\), we obtain the components
	\begin{subequations}
		\begin{align}
			\mathcal{A}^r &= \frac{1}{\Xi^2}\left( r'' - \frac{\Xi'}{\Xi} r' + \frac{1}{2\alpha_{rr}}\frac{d\alpha_{rr}}{dr} (r')^2 - \frac{1}{2\alpha_{rr}}\frac{d\alpha_{\phi\phi}}{dr} \right), \\
			\mathcal{A}^\phi &= \frac{1}{\Xi^2}\left( -\frac{\Xi'}{\Xi} + \frac{1}{\alpha_{\phi\phi}}\frac{d\alpha_{\phi\phi}}{dr} r' \right).
		\end{align}
	\end{subequations}
	
	Inserting these into the geodesic curvature formula~\eqref{geocur} and simplifying, we arrive at the general expression
	\begin{align}
		\label{geocub}
		k_g =& \frac{\sqrt{\alpha}}{\Xi^2}\bigg[ -r'' + \frac{1}{\alpha_{\phi\phi}}\frac{d\alpha_{\phi\phi}}{dr} (r')^2 - \frac{1}{2\alpha_{rr}}\frac{d\alpha_{rr}}{dr} (r')^2 \notag\\
		&+ \frac{1}{2\alpha_{rr}}\frac{d\alpha_{\phi\phi}}{dr} \bigg]\frac{d\phi}{dt}. 
	\end{align}
	This is the geodesic curvature of a general curve \(r = r(\phi)\). For the straight line \(r = b/\sin\phi\) considered in this work, substituting
	\begin{align}
		\label{XiaLGH}
	r' = -\frac{b\cos\phi}{\sin^2\phi},\quad r'' = \frac{b(2\cos^2\phi + \sin^2\phi)}{\sin^3\phi}
\end{align}
	into \eqref{geocub} and expanding the metric coefficients in the weak‑field limit yields the leading‑order geodesic curvature. Substituting this result into Eq.~\eqref{leadefangexp} then gives the deflection angle to leading order.

\subsection{Stationary spacetime}
\label{WLuiAI}
For stationary spacetimes, the optical geometry is no longer Riemannian but is described by a Randers-type Finsler metric of the form~\cite{Randers,Cheng-Shen} 
\begin{align}
F(x,y) = \sqrt{\alpha_{ij}(x)\,y^i y^j} + \beta_i(x)\,y^i,
\end{align}
where $x\in\mathcal{M}$ is a point on the manifold and $y\in T_x\mathcal{M}$ a tangent vector at $x$. Here $\alpha_{ij}(x)$ is a Riemannian metric on $\mathcal{M}$, and $\beta_i(x)$ is a one-form.

By adopting Werner’s method~\cite{Werner2012}, namely employing Nazım’s construction to obtain 
the osculating Riemannian metric of a Randers space---which by construction shares the same 
geodesics as the Randers metric---and thereby applying the Gauss--Bonnet theorem to light 
deflection, the deflection angle in Eq.~\eqref{Gibbons-Werner} remains valid. The osculating Riemannian metric is obtained by choosing a smooth nonvanishing tangent vector field $V(x)$ along the geodesic, 
\begin{align}
	\bar g_{ij}(x) = g_{ij}(x,V(x)),
\end{align}
where the fundamental tensor of $F$ is defined as
\begin{align}
g_{ij}(x,y) = \tfrac{1}{2}\,\frac{\partial^2 F^2}{\partial y^i \partial y^j}(x,y).
\end{align}

However, implementing Werner’s method in the $(r,\phi)$ coordinates involves considerable computational complexity. To simplify the analysis, Ref.~\cite{Lizz_PRD2025} proposed carrying out the calculation in the Cartesian-like coordinates $(X,Y)$ with $X=r\cos\phi$ and $Y=r\sin\phi$. In these coordinates, the trajectory of the curve $\gamma_b$ is simply given by $Y=b$, and one may choose $V^i = (-1,0)$ to construct the osculating Riemannian metric~\cite{Lizz_PRD2025}. Accordingly, in the $(X,Y)$ coordinates the optical osculating Riemannian metric takes the form
\begin{align}
	dt^2 = \bar{g}_{ij}\,dx^i dx^j 
	= \bar{g}_{XX}\, dX^2 + 2 \bar{g}_{XY}\, dX\, dY + \bar{g}_{YY}\, dY^2.\nn
\end{align}

\begin{figure}[!t]
	\centering
	\includegraphics[width=8cm]{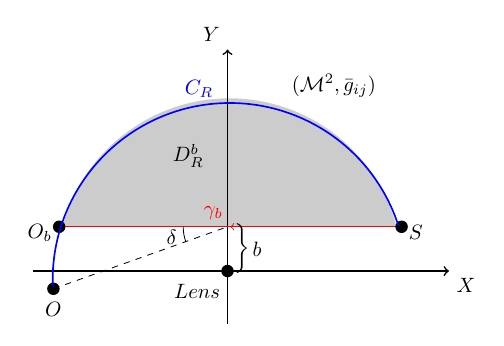}
	\caption{Region $D_R^b \subset (\mathcal{M}^{2},\bar{g}_{ij})$ with coordinates $(X,Y)$. 
		The straight line $\gamma_b$ is described by $Y=b$. 
		Note that the coordinates $X$ and $Y$ are not necessarily orthogonal.}
	\label{XuWei_XY}
\end{figure}

In the two-dimensional optical osculating Riemannian space $(\mathcal{M}^2,\bar{g}_{ij})$, 
the discussion of Sec.~\ref{Czesław Miłosz} remains valid, and the deflection angle in 
Eq.~\eqref{Gary Snyder} can be expanded as (see Fig.~\ref{XuWei_XY})
\begin{align}
	\label{leadefangXY}
	{}^{L}\delta 
	&= -\int_{S}^{O_b} {}^{L}k_g(\gamma_b)\, dt \nn\\
	&= \int_{-\infty}^{\infty} {}^{L}\!\left[k_g(\gamma_b)\,\frac{dt}{dX}\right]_{Y=b}\, dX .
\end{align}

For the straight line $\gamma_b$ defined by $Y=b$, the unit tangent vector is
\begin{align}
	\label{TanXY}
	\mathcal{T}^i = \left(\frac{dX}{dt},\,\frac{dY}{dt}\right) 
	= \frac{1}{\sqrt{\bar{g}_{XX}}}(1,0).
\end{align}
Along $\gamma_b$, the unit normal vector $N^i$ orthogonal to $\mathcal{T}^i$ satisfies
\begin{align}
	\bar{g}_{ij}N^i N^j = 1, 
	\qquad \bar{g}_{ij}N^i \mathcal{T}^j = 0,
\end{align}
and can be chosen such that $\{\mathcal{T}^i,N^i\}$ forms a positively oriented orthonormal frame:
\begin{align}
	\label{norvelXY}
	N^i = \frac{1}{\sqrt{\bar{g}}}\left(-\frac{\bar{g}_{XY}}{\sqrt{\bar{g}_{XX}}},\, \sqrt{\bar{g}_{XX}}\right),
\end{align}
where $\bar{g} = \det(\bar{g}_{ij}) = \bar{g}_{XX}\bar{g}_{YY} - \bar{g}_{XY}^2$.

The geodesic curvature of $\gamma_b$ is then
\begin{align}
	\label{geodXY}
	k_g(\gamma_b) = \bar{g}_{ij} N^i \mathcal{A}^j \Big|_{Y=b},
\end{align}
where the acceleration vector is defined as
\begin{align}
	\label{AccXY}
	\mathcal{A}^j = \nabla_{\mathcal{T}} \mathcal{T}^j 
	= \frac{d\mathcal{T}^j}{dt} + \bar\Gamma^j_{kl}\,\mathcal{T}^k \mathcal{T}^l,
\end{align}
with $\bar\Gamma^j_{kl}$ denoting the Christoffel symbols associated with the
two-dimensional metric $\bar g_{ij}$.

Substituting Eq.~\eqref{norvelXY} into Eq.~\eqref{geodXY}, the geodesic curvature is obtained as
\begin{align}
	\label{XYgeode}
	k_g(\gamma_b) 
	= \frac{\sqrt{\bar{g}}}{\sqrt{\bar{g}_{XX}}}\,\mathcal{A}^Y \Big|_{Y=b} = \frac{\sqrt{\bar{g}}}{\bar{g}_{XX}}\,\bar\Gamma_{XX}^Y\,\Big|_{Y=b} \frac{dX}{dt},
\end{align}
where the relations \eqref{TanXY} and \eqref{AccXY} have been used. Substituting the above into Eq.~\eqref{leadefangXY}, we obtain the deflection angle
\begin{align}
	\label{defXYang}
	{}^{L}\delta=\int_{-\infty}^{\infty}{}^{L}\left[\frac{\sqrt{\bar{g}}}{\bar{g}_{XX}}~\bar{\Gamma}_{XX}^Y\right]\Big|_{Y=b}~dX.
\end{align}

\section{Applications: light deflection in typical spacetimes}
\label{Patrick Modiano}

In this section, we apply our method to compute the leading-order bending angle of light in curved spacetimes. We focus on the three most famous cases: the Schwarzschild, Reissner-Nordstr\"om, and Kerr spacetimes.

\subsection{Schwarzschild spacetime}

The line element of Schwarzschild spacetime is  
\begin{align}  
	ds^2 = &-\left(1 - \frac{2M}{r}\right) \, dt^2 + \frac{1}{1 - \frac{2M}{r}} \, dr^2 \nn\\
	&+ r^2 \left( d\theta^2 + \sin^2\theta \, d\phi^2 \right),  
\end{align}  
where \( M \) denotes the mass of the central body. 

Restricting to the equatorial plane ($\theta = \pi/2$), the null condition \(ds^2 = 0\) yields the optical metric
	\begin{align}
		\label{Schoptmet}
		dt^2 = \frac{dr^2}{\left(1 - \frac{2M}{r}\right)^2} + \frac{r^2}{1 - \frac{2M}{r}} d\phi^2,
	\end{align}
	with components
	\begin{align}
		\alpha_{rr} = \frac{1}{\left(1 - \frac{2M}{r}\right)^2}, \qquad
		\alpha_{\phi\phi} = \frac{r^2}{1 - \frac{2M}{r}},
	\end{align}
	and determinant
	\begin{align}
	\alpha =\alpha_{rr}\alpha_{\phi\phi}=\frac{r^2}{\left(1 - \frac{2M}{r}\right)^3}.
	\end{align}

Substituting these into the general expression \eqref{geocub} gives the geodesic curvature for an arbitrary curve \(r = r(\phi)\) in the Schwarzschild optical geometry:
\begin{align}
	{}^{L}k_g =&\bigg\{ \frac{r}{(r')^2 + r^2} \Bigg[ -r'' + \frac{2}{r}(r')^2 + r\bigg]\nn\\
	& + \frac{M}{\left((r')^2 + r^2\right)^2}\bigg[  (r')^2(r''-2r) \nn\\
	&-r^2(r''+2r)- \frac{2}{r}(r')^4 \bigg]\Bigg\}\frac{d\phi}{dt}.
\end{align}

For the straight line \(r = b/\sin\phi\), substituting the derivatives \(r'\) and \(r''\) from \eqref{XiaLGH} into the above expression yields the leading-order geodesic curvature
\begin{align}
	{}^{L}k_g(\gamma_b)=-\frac{3M\sin^3\phi}{b}\frac{d\phi}{dt}.
\end{align}

Substituting it into Eq.~\eqref{leadefangexp}, we obtain the deflection angle as
\begin{align}
	{}^{L}\delta 
	&= - \int_{0}^{\pi} {}^{L}\!\left[ k_g(\gamma_b)\,\frac{dt}{d\phi} \right] d\phi \nn\\
	&= \frac{3M}{b} \int_{0}^{\pi} \sin^{3}\!\phi \, d\phi \nn\\
	&= \frac{4M}{b}.
\end{align}
which reproduces the classical weak-field result for light deflection.

\subsection{Reissner--Nordstr\"om spacetime}

The line element of the Reissner--Nordstr\"om spacetime reads
\begin{align}
	\label{RNmelsp}
	ds^2 
	=& -\left(1 - \frac{2M}{r}+\frac{Q^2}{r^2}\right)\mathrm{d}t^2
	+ \frac{1}{1 - \frac{2M}{r}+\frac{Q^2}{r^2}}\mathrm{d}r^2 \nn\\
	&+ r^2\left(\mathrm{d}\theta^2 + \sin^2\theta\,\mathrm{d}\phi^2\right),
\end{align}
where $M$ denotes the mass and $Q$ the electric charge of the central body.  

	On the equatorial plane ($\theta=\pi/2$), the optical metric takes the form
	\begin{align}
		\label{RNoptmet}
		\mathrm{d}t^2 
		= \frac{\mathrm{d}r^2}{\left(1 - \frac{2M}{r}+\frac{Q^2}{r^2}\right)^2}
		+ \frac{r^2}{1 - \frac{2M}{r}+\frac{Q^2}{r^2}}\,\mathrm{d}\phi^2,
	\end{align}
	with the metric components and determinant
	\begin{subequations}
		\label{JinRWu}
		\begin{align}
			\alpha_{rr} &= \frac{1}{\left(1 - \frac{2M}{r}+\frac{Q^2}{r^2}\right)^2}, \quad
			\alpha_{\phi\phi} = \frac{r^2}{1 - \frac{2M}{r}+\frac{Q^2}{r^2}}, \\
			\alpha &= \alpha_{rr}\alpha_{\phi\phi} = \frac{r^2}{\left(1 - \frac{2M}{r}+\frac{Q^2}{r^2}\right)^3}.
		\end{align}
	\end{subequations}

Now, following the same approach as for the Schwarzschild case, we compute the geodesic curvature of the straight line $\gamma_{b}$ in the Reissner--Nordstr\"om optical geometry. Substituting the metric quantities given above into the general expression \eqref{geocub} and specializing to the straight line \(r = b/\sin\phi\) (with \(r'\) and \(r''\) as in Eq.~\eqref{XiaLGH}), we expand to leading order in \(M\) and \(Q^2\). This yields the leading-order geodesic curvature
\begin{align}
	{}^{L}k_g(\gamma_b)= \left(-\frac{3M\sin^3\phi}{b} + \frac{Q^2\sin^4\phi}{b^2}\right)\frac{d\phi}{dt}.
\end{align}
Inserting this into Eq.~\eqref{leadefangexp}, the deflection angle becomes
\begin{align}
	\label{RNphlzh}
	{}^{L}\delta 
	&= \frac{3M}{b}\int_{0}^{\pi} \sin^3\phi \, d\phi
	- \frac{Q^2}{b^2}\int_{0}^{\pi} \sin^4\phi \, d\phi \nn\\
	&= \frac{4M}{b} - \frac{3\pi Q^2}{4b^2},
\end{align}
which reproduces the known weak-field result (see, e.g., Refs.~\cite{He_Lin_CQG2016,Pang_Jia}).

\subsection{Kerr spacetime}

The Kerr metric describes the spacetime outside a rotating body of mass $M$ with 
specific angular momentum $a$. In Boyer--Lindquist coordinates $(t, r, \theta, \phi)$, 
the line element is given by~\cite{Kerr-BL}
\begin{align}
	ds^2 = &-\left(1-\frac{2Mr}{\Sigma}\right)dt^2 
	+ \frac{\Sigma}{\Delta}dr^2 + \Sigma\, d\theta^2 \nn\\
	&+ \frac{1}{\Sigma}\Big[\,(r^2+a^2)^2 - \Delta a^2\sin^2\theta\,\Big]\sin^2\theta\, d\phi^2 \nn\\
	&- \frac{4Mar}{\Sigma}\sin^2\theta\, dt\, d\phi,
\end{align}
where
\[
\Sigma = r^2 + a^2\cos^2\theta, 
\qquad 
\Delta = r^2 - 2Mr + a^2.
\]

Using the null condition $ds^2=0$ and restricting to the equatorial plane 
$(\theta=\pi/2)$, the Kerr optical Randers metric takes the form~\cite{Werner2012}
\begin{align}
	dt = \sqrt{\alpha_{ij}\,dx^i dx^j} + \beta_i\, dx^i,
\end{align}
where
\begin{subequations}
	\small
	\begin{align}
		\alpha_{ij}dx^i dx^j &= \frac{r^2}{\Delta-a^2}\left(\frac{r^2}{\Delta}\,dr^2 
		+ \frac{r^2\Delta}{\Delta-a^2}\, d\phi^2\right),\\
		\beta_i dx^i &= -\frac{2Mar}{\Delta-a^2}\, d\phi.
	\end{align}
\end{subequations}

For computational simplicity, we adopt the osculating Riemannian metric associated with the above Randers form in the $(X,Y)$ coordinates. Since deriving it is not the focus of this work, we directly quote the result from Eq.~(28) of Ref.~\cite{Lizz_PRD2025}. As Ref.~\cite{Lizz_PRD2025} treats the motion of massive particles, we set $E=1$ and $v=1$ therein, yielding
\begin{subequations}
	\begin{align}
		\label{KerrXX}
		{}^{L}\bar{g}_{XX} &= 1 + \frac{2M}{r^3}\bigl(r^2 + X^2 - 2aY\bigr),\\
		{}^{L}\bar{g}_{XY} &= \frac{2MX}{r^3}(Y+a),\\
		{}^{L}\bar{g}_{YY} &= 1 + \frac{2M}{r^3}\bigl(r^2 + Y^2 - aY\bigr),
	\end{align}
\end{subequations}
where $r=\sqrt{X^2+Y^2}$. The determinant of the osculating metric is then
	\begin{align}
		\label{KerrHang}
	{}^{L}\bar{g} = 1 + \frac{6M}{r}\left(1-\frac{aY}{r^2}\right) .
 \end{align}

From the osculating metric $\bar{g}_{ij}$, the relevant Christoffel symbol is
\begin{align}
	\label{Tomas Tranströmer}
	{}^{L}\bar{\Gamma}^{Y}{}_{XX}\big|_{Y=b}
	= \frac{M\left(3b^{3}-2a\left(X^{2}+b^{2}\right)\right)}{(X^{2}+b^{2})^{5/2}} .
\end{align}
Since the Christoffel symbol is at least of order $M$, the geodesic curvature formula~\eqref{XYgeode} only requires the zeroth-order term of $\sqrt{\bar g}/\bar{g}_{XX}$, which from Eqs.~\eqref{KerrXX} and~\eqref{KerrHang} is $1$. This leads to the following expression for the geodesic curvature:
\begin{align}
	{}^{L}k_g(\gamma_b)
	= \frac{M\left(3b^{3}-2a\left(X^{2}+b^{2}\right)\right)}{(X^{2}+b^{2})^{5/2}}\,\frac{dX}{dt}.
\end{align}
Substituting this into Eq.~\eqref{leadefangXY}, equivalently substituting Eq.~\eqref{Tomas Tranströmer} into Eq.~\eqref{defXYang}, the deflection angle is found to be
\begin{align}
	{}^{L}\delta=&\int_{-\infty}^{\infty}{}^{L}\left[k_g(\gamma_b)\frac{dt}{dX}\right]dX\nn\\
	=&\int_{-\infty}^{\infty}{}^{L}\frac{M\left[3b^{3}-2a\left(X^{2}+b^{2}\right)\right]}{\left(X^{2}+b^{2}\right)^{5/2}}dX\nn\\
	=&\frac{4M}{b}-\frac{4Ma}{b^2}.
\end{align}
The above expression corresponds to prograde light trajectories. The retrograde case is obtained by replacing $a\to-a$~\cite{Lizz_PRD2025}. Combining both cases, the leading-order deflection angle reads~\cite{Edery_GodinPRD2025,Sereno_De Luca,Werner2012}
\begin{align}
	\label{DA-Kerr}
	{}^{L}\delta = \frac{4M}{b} \pm \frac{4Ma}{b^2},
\end{align}
where the upper (lower) sign refers to retrograde (prograde) trajectories. 

	\section{Extension to massive particles}
	\label{AIQING}

 Having applied the newly established formula \eqref{Gary Snyder} to typical spacetimes and obtained the leading-order weak-field deflection angle for light, we now extend the method to massive particles via the Jacobi metric~\cite{Gibbons2016,Chanda2019}. Just as light rays are geodesics of the optical metric, trajectories of massive particles can be regarded as geodesics of the Jacobi metric. To emphasize the physical intuition while avoiding unnecessary computational complexity, we focus on the deflection of neutral massive particles in static spacetimes, taking the Reissner–Nordstr\"om case as a concrete example.
	
	\subsection{Jacobi metric for static spacetimes}
	
	For a static, spherically symmetric spacetime given by Eq.~\eqref{ssspacetime}, the corresponding Jacobi metric for a massive particle of mass \(m\) and energy \(E\) is given by~\cite{Gibbons2016}
	\begin{align}
		\label{Jacobi_metric}
		d\ell^2 =J_{ij}dx^idx^j =\left[E^2 - m^2 A(r)\right]\alpha_{ij}dx^idx^j,
	\end{align}
	where $\alpha_{ij}$ is the optical metric given by Eq.~\eqref{layonha}.
	
The energy is related to the asymptotic velocity \(v\) by
\begin{align}
	E = \frac{m}{\sqrt{1-v^2}}.
\end{align}
	
Using this relation and considering the equatorial plane (\(\theta=\pi/2\)), the Jacobi metric \eqref{Jacobi_metric} becomes
\begin{align}
	\label{epjacobi}
	d\ell^2 &= J_{ij}dx^i dx^j \nn\\
	= &m^2\left[\frac{1}{A(1-v^2)} - 1\right]\left[Bdr^2 + r^2 d\phi^2\right].
\end{align}

	\subsection{Deflection formula for massive particle}
	
	The extension to massive particles is completely analogous to the photon case: one simply replaces the optical metric $\alpha_{ij}$ in the two-dimensional spatial space by the Jacobi metric $J_{ij}$. Under the same assumptions---asymptotically flat space, weak-field and small-angle approximation, source and observer at infinity, and leading-order treatment---the central result of this work, Eq.~\eqref{Gary Snyder}, becomes
	\begin{align}
		{}^{L}\delta = -\int_{S}^{O_b} {}^{L}k_g(\gamma_b)\, d\ell,
	\end{align}
where $d\ell$ is the arc element of the Jacobi metric and $k_g(\gamma_b)$ is the geodesic curvature of the straight line $\gamma_{b}$ (with $r=b/\sin\phi$) with respect to $J_{ij}$. Expressed in terms of the angular coordinate, the deflection angle takes the form
	\begin{align}
		\label{LmassDef}
		{}^{L}\delta = -\int_{0}^{\pi} {}^{L}\!\left[k_g(\gamma_b)\left(\frac{d\ell}{d\phi}\right)\right] d\phi .
	\end{align}
For massive particles, the geodesic curvature formula \eqref{geocub} is replaced by its Jacobi metric analogue
\begin{align}
	\label{massgeocub}
	k_g =& \frac{\sqrt{J}}{\Xi_J^2}\bigg[ -r'' + \frac{1}{J_{\phi\phi}}\frac{dJ_{\phi\phi}}{dr} (r')^2 - \frac{1}{2J_{rr}}\frac{dJ_{rr}}{dr} (r')^2 \notag\\
	&+ \frac{1}{2J_{rr}}\frac{dJ_{\phi\phi}}{dr} \bigg]\frac{d\phi}{d\ell},
\end{align}
where
\begin{align}
&J = \det(J_{ij}) = J_{rr}J_{\phi\phi},\quad 
\Xi_J = \sqrt{J_{rr}(r')^2 + J_{\phi\phi}},\nn\\
&r' = \frac{dr}{d\phi},\quad 
r'' = \frac{d^2r}{d\phi^2}.
\end{align}

\subsection{Application to Reissner--Nordstr\"om spacetime}
	
We now apply the method to a massive particle moving in the equatorial plane of the Reissner--Nordstr\"om spacetime, whose line element is given in Eq.~\eqref{RNmelsp}. Substituting the metric functions into Eq.~\eqref{epjacobi} and expanding to leading order in the weak-field approximation yields the equatorial plane Jacobi metric components
\begin{subequations}
	\begin{align}
		{}^{L}J_{rr} &= \frac{m^2}{1 - v^2} \left[ v^2+\left(1 + v^2\right)\left( \frac{2M}{r} - \frac{Q^2}{r^2}\right) \right], \\
		{}^{L}J_{\phi\phi} &= \frac{m^2}{1 - v^2} \left( r^2 v^2 + 2Mr - Q^2 \right).
	\end{align}
\end{subequations}
The determinant is given by
\begin{align}
	{}^{L}J = \frac{m^4 v^2}{\left(1 - v^2\right)^2} \bigg[ r^2 v^2 + \left(2 + v^2\right) (2Mr- Q^2) \bigg].
\end{align}

Substituting these quantities into Eq.~\eqref{massgeocub}, specializing to the straight line \(r = b/\sin\phi\) (with \(r',r''\) from Eq.~\eqref{XiaLGH}), and expanding to leading order, we obtain the geodesic curvature
	\begin{align}
		{}^{L}k_g(\gamma_b) =& \bigg[ -\left(2 + v^2 - 3v^2 \cos 2\phi\right) \frac{M \sin\phi}{2b v^2}\nn\\
		& + \left(1 - v^2 \cos 2\phi\right) \frac{Q^2 \sin^2\phi}{b^2 v^2} \bigg]\frac{d\phi}{d\ell}.
	\end{align}
Inserting this into Eq.~\eqref{LmassDef}, the deflection angle becomes
	\begin{align}
		{}^{L}\delta &= \frac{2M}{b}\left(1+\frac{1}{v^2}\right) - \frac{\pi Q^2}{4b^2}\left(1+\frac{2}{v^2}\right).
	\end{align}
This result is consistent with previous work (see, e.g., Refs.~\cite{He_Lin_CQG2016,Pang_Jia}). In the limit \(v\to 1\), the above expression reduces to the light deflection angle \eqref{RNphlzh}.
	
\subsection{Further generalizations}

For stationary, axisymmetric spacetimes, the Jacobi metric takes the form of a Randers–Finsler metric, completely analogous to the optical Randers case. Following the same procedure as in Sec.~\ref{WLuiAI}, we apply Werner's method~\cite{Werner2012} and perform the calculation in Cartesian-like coordinates~\cite{Lizz_PRD2025}. In this case, the deflection angle formula retains the form of \eqref{Gary Snyder} as well as its more detailed expression \eqref{defXYang}, with the optical metric simply replaced by the corresponding Jacobi metric.

In addition, the deflection of charged particles in combined gravitational and electromagnetic fields is also worth noting. Depending on the specific model, the Jacobi metric may be Riemannian (e.g., describing a charged particle in the Reissner–Nordström spacetime~\cite{Crisnejo2019}) or of Randers type (e.g., describing a charged particle in a Schwarzschild spacetime with a dipole magnetic field~\cite{massiveGB-LiWJ}). In either case, the treatment is essentially the same as for neutral particles in static or stationary spacetimes, respectively.

\section{Conclusion}
\label{Marguerite Yourcenar}
	In this work, we revisit the problem of weak-field gravitational deflection from a new perspective. Instead of following the conventional approach of tracking the change in the coordinate angle, we employ the Gauss–Bonnet theorem to show that, to leading order, the deflection angle of light can be expressed as the integral of the geodesic curvature of a straight line in curved optical space (see Eq.~\eqref{Gary Snyder}). We apply this new formula to three well-known spacetimes—Schwarzschild, Reissner–Nordstr\"om, and Kerr—and recover the standard leading-order results in each case. Beyond null geodesics, we extend the framework to massive particles via the Jacobi metric, and explicitly compute the deflection angle for massive particles in the Reissner–Nordstr\"om spacetime, obtaining velocity-dependent corrections.
	
	Our aim is not merely to reproduce known results, but to emphasize the conceptual shift underlying this reformulation. This formulation reveals a geometric duality: gravitational deflection can be understood either as the bending of a curve in flat space or as the bending of a straight line in curved space. It clearly demonstrates that the deflection angle is a global, coordinate-independent quantity and provides a complementary viewpoint to the classical Gibbons–Werner method.
	
	The method presented here can be extended in several directions. For the test particle, additional parameters such as spin can be incorporated besides electric charge. For the lens system, the framework can be generalized to asymptotically non‑flat backgrounds, modified theories of gravity, or environments such as plasma. Exploring finite‑distance corrections and higher‑order deflection angles are also worthwhile avenues for future investigation. Finally, from an observational perspective, the viewpoint introduced here may offer a new way to probe gravity: by manipulating particles to move along a prescribed straight line and comparing the difference in energy loss required for such motion in the presence versus the absence of a gravitational field (such as that of the Sun), one could potentially extract information about the gravitational body.
	
\acknowledgements
This work was supported by the National Natural Science Foundation of China under Grant No.12565009.


\begin{thebibliography}{99}
	
\bibitem{Amari-info-geo} S. Amari, {\it Information Geometry and Its Applications} (Springer, Tokyo, 2016).
\bibitem{Weinhold-1975} F. Weinhold, J. Chem. Phys. {\bf 63}(6), 2479 (1975).
\bibitem{Ruppeiner-1995} G. Ruppeiner, Rev. Mod. Phys. {\bf 67}(3), 605 (1995).
\bibitem{Ingarden-1979} R. S. Ingarden, Y. Sato, K. Sugawa, and M. Kawaguchi, Tensor (N.S.) {\bf 33}, 347 (1979).
\bibitem{Janyszek-1990} H. Janyszek, J. Phys. A: Math. Gen. {\bf 23}(4), 477 (1990).
\bibitem{Ruppeiner2014} G.~Ruppeiner, Springer Proc.\ Phys.\ {\bf 153}, 179 (2014).
\bibitem{Wei-2019} S. -W. Wei, Y. -X. Liu, and R. B. Mann, Phys. Rev. D {\bf 100}, 124033 (2019).
\bibitem{Antonelli-2003} P. L. Antonelli (Ed.), {\it Handbook of Finsler Geometry}, Vol.~1 (Kluwer Academic, Dordrecht, 2003).
\bibitem{Berry-1984} M. V. Berry, Proc. R. Soc. A {\bf 392}, 45 (1984).
\bibitem{Nielsen-2006-QIC} M. A. Nielsen, Quantum Inf. Comput. {\bf 6}, 213 (2006).
\bibitem{Dowling-2008-QIC} M. R. Dowling and M. A. Nielsen, Quantum Inf. Comput. {\bf 8}(10), 861 (2008).
\bibitem{Jefferson-2017-JHEP} R. A. Jefferson and R. C. Myers, J. High Energy Phys. {\bf 2017}(10), 1 (2017).
\bibitem{Susskind-2020} L. Susskind, {\it Three Lectures on Complexity and Black Holes} (Springer, Berlin/Heidelberg, 2020).
\bibitem{Liu-2011-PRA} Q. H. Liu, L. H. Tang, and D. M. Xun, Phys. Rev. A {\bf 84}(4), 042101 (2011).
\bibitem{Russell-2014} B. Russell, S. Stepney, Phys. Rev. A {\bf 90}, 012303 (2014).  
\bibitem{Brody-2015-PRL} D. C. Brody, D. M. Meier, Phys. Rev. Lett. {\bf 114}, 100502 (2015).  
\bibitem{Arnold-mech} V. I. Arnold, {\it Mathematical Methods of Classical Mechanics}, 2nd ed. (Springer, New York, 1989).
\bibitem{Awrejcewicz-mech} J. Awrejcewicz, {\it Classical Mechanics: Dynamics} (Springer, New York, 2012).
\bibitem{Weyl} H. Weyl, Ann. Phys. (Berlin) {\bf359}, 117 (1917).
\bibitem{Perlick} V. Perlick, {\it Ray Optics, Fermat’s Principle, and Applications to General Relativity} (Springer, New York, 2000).
\bibitem{Gibbons2016} G. W. Gibbons, Classical Quantum Gravity {\bf33}, 025004(2016).
\bibitem{Chanda2019} S. Chanda, G. W. Gibbons, P. Guha, P. Maraner, and M. C. Werner, J. Math. Phys. (N.Y.) {\bf60}, 122501 (2019).
\bibitem{Das-EPJC-2017} P. Das, R. Sk. and S. Ghosh, Eur. Phys. J. C {\bf 77}, 735 (2017).
\bibitem{Arganaraz-CQG-2021} M. Arga{\~n}araz and O. Lasso Andino, Class. Quant. Grav. {\bf 38}, 045004 (2021).
\bibitem{Qiao-PRD-2022a} C.-K. Qiao and M. Li, Phys. Rev. D {\bf 106}, 021501 (2022).  
\bibitem{Qiao-PRD-2022b} C.-K. Qiao, Phys. Rev. D {\bf 106}, 084060 (2022).  
\bibitem{Qiao-EPJC-2025} C.-K. Qiao, Eur. Phys. J. C {\bf 85}, 191 (2025).
\bibitem{Cunha-CQG-2022} P. V. P. Cunha, C. A. R. Herdeiro and J. P. A. Novo, Class. Quant. Grav. {\bf 39}, 225007 (2022).  
\bibitem{Bermudez-2025} B. Berm\'udez-C\'ardenas and O. L. Andino, arXiv:2503.21203.  
\bibitem{Congdon-Keeton-2018} A. B. Congdon and C. R. Keeton, {\it Principles of Gravitational Lensing} (Springer International Publishing, Cham, 2018).
\bibitem{Gibbons-Werner} G. W. Gibbons and M. C. Werner, Classical Quantum Gravity {\bf 25}, 235009 (2008).
\bibitem{Werner2012} M. C. Werner, Gen. Relativ. Gravit. {\bf 44}, 3047 (2012).
\bibitem{OIA2017} T. Ono, A. Ishihara, and H. Asada, Phys. Rev. D {\bf 96}, 104037 (2017).
\bibitem{massiveGB-CG} G. Crisnejo and E. Gallo, Phys. Rev. D {\bf 97}, 124016 (2018).
\bibitem{massiveGB-CGJ} G. Crisnejo, E. Gallo, and K. Jusufi, Phys. Rev. D {\bf 100}, 104045 (2019).
\bibitem{Crisnejo2019} G. Crisnejo, E. Gallo, and J. R. Villanueva, Phys. Rev. D {\bf 100}, 044006 (2019).
\bibitem{massiveGB-LiHZ} Z. Li, G. He, and T. Zhou, Phys. Rev. D {\bf 101}, 044001 (2020).
\bibitem{massiveGB-LiJa2020} Z. Li and J. Jia, Eur. Phys. J. C {\bf80}, 157 (2020).
\bibitem{massiveGB-LiWJ} Z. Li, W. Wang, and J. Jia, Phys. Rev. D {\bf106}, 124025 (2022).
\bibitem{LiJia-2024} Z. Li and J. Jia, Eur. Phys. J. C {\bf84}, 989 (2024).
\bibitem{ISOA2016} A. Ishihara, Y. Suzuki, T. Ono, T. Kitamura, and H. Asada, Phys. Rev. D {\bf 94}, 084015 (2016).
\bibitem{OIA-OA} T. Ono and H. Asada, Universe, {\bf5}, 218 (2019).
\bibitem{TOA} K. Takizawa, T. Ono, and H. Asada, Phys. Rev. D {\bf101}, 104032 (2020).
\bibitem{massiveGB-LiZA} Z. Li, G. Zhang and A.\"{O}vg\"{u}n, Phys. Rev. D {\bf101}, 124058 (2020).
\bibitem{HuangCa} Y. Huang and Z. Cao, Eur. Phys. J. C {\bf83}, 80 (2023).
\bibitem{HuangCb} Y. Huang and Z. Cao, Phys. Rev. D {\bf106}, 104043(2022).
\bibitem{HuangCL} Y. Huang, Z. Cao and Z. Lu, J. Cosmol. Astropart. Phys. 01 (2024) 013.
\bibitem{HuangSC} Y. Huang, B. Sun, and Z. Cao, Phys.Rev.D {\bf107}, 104046 (2023).
\bibitem{Jusufi-Ovgun-2018} K. Jusufi and A. \"{O}vg\"{u}n, Phys. Rev. D {\bf97}, 064030 (2018).  
\bibitem{Ono-Ishihara-Asada-2018} T. Ono, A. Ishihara and H. Asada, Phys. Rev. D {\bf98}, 044047 (2018).  
\bibitem{Ovgun-2018} A. \"{O}vg\"{u}n, Phys. Rev. D {\bf98}, 044033 (2018).  
\bibitem{Ovgun-Sakalli-Saavedra-2018} A. \"{O}vg\"{u}n, I. Sakall{\i} and J. Saavedra, J. Cosmol. Astropart. Phys. {\bf10}, 041 (2018).  
\bibitem{Ovgun-Sakalli-Saavedra-2019} A. \"{O}vg\"{u}n, I. Sakall{\i} and J. Saavedra, Annals Phys. {\bf411}, 167978 (2019).  
\bibitem{Crisnejo-Gallo-Rogers-2019} G. Crisnejo, E. Gallo and A. Rogers, Phys. Rev. D {\bf99}, 124001 (2019).  
\bibitem{Gao:2023ltr} 
X.-J.~Gao, X.-K.~Yan, Y. Yin and Y.-P.~Hu, Eur. Phys. J. C {\bf 83}, 281 (2023).
\bibitem{Ali2025a} R. Ali, X. Tiecheng, M. Awais, and R. Babar, Ann. Phys. (N.Y.) {\bf 482}, 170201 (2025).
\bibitem{Ali2024b} R. Ali, X. Tiecheng, M. Awais, and R. Babar, Commun. Theor. Phys. {\bf 76}, 095404 (2024).
\bibitem{Gibbons-PRD-2009a} G. W. Gibbons and C. M. Warnick, Phys. Rev. D {\bf 79}, 064031 (2009).  
\bibitem{Gibbons-PRD-2009b} G. W. Gibbons, C. A. R. Herdeiro, C. M. Warnick and M. C. Werner, Phys. Rev. D {\bf 79}, 044022 (2009).  
\bibitem{Petters-Werner-GRG-2010} A. O. Petters and M. C. Werner, Gen. Relativ. Gravit. {\bf 42}, 2011–2046 (2010).  
\bibitem{Werner-Obs-2010} M. C. Werner, The Observatory {\bf 130}, 193–194 (2010).  
\bibitem{Roesch-Werner-PAMQ-2020} H. P. Roesch and M. C. Werner, Pure Appl. Math. Q. {\bf 16}, 495–514 (2020).  
\bibitem{Halla-thesis-2022} M. Halla, {\it Investigation of the Gravitational Lens Effect with Differential Topology}, Ph.D. Thesis, Universität Bremen (2022).  
\bibitem{Li-Zhou} Z. Li and T. Zhou, Phys. Rev. D {\bf101}, 044043 (2020).
\bibitem{doCarmo} M. P. do Carmo, {\it Differential Geometry of Curves and Surfaces} (Prentice Hall, Englewood Cliffs, 1976).
\bibitem{Tu-diff-geo} L. W. Tu, {\it Differential Geometry: Connections, Curvature, and Characteristic Classes} (Springer, Cham, 2017).
\bibitem{Randers} G. Randers, Phys. Rev. {\bf59}, 195–199 (1941).
\bibitem{Cheng-Shen} X. Cheng and Z. Shen, {\it Finsler Geometry, An Approach via Randers Spaces} (Springer-Verlag, Berlin, 2012).
 \bibitem{Lizz_PRD2025} Z. Li, Phys. Rev. D {\bf111}, 084017 (2025).
 \bibitem{He_Lin_CQG2016} G. He and W. Lin, Classical Quantum Gravity {\bf33}, 095007 (2016).
 \bibitem{Pang_Jia}  X. Pang and J. Jia, Classical Quantum Gravity {\bf36}, 065012 (2019).
 \bibitem{Kerr-BL}  R. H. Boyer, R. W. Lindquist, J. Math. Phys. (N.Y.) {\bf 8}, 265 (1967).
 \bibitem{Edery_GodinPRD2025} A. Edery and J. Godin, Gen. Relativ. Gravit. {\bf 38}, 1715 (2006).
 \bibitem{Sereno_De Luca} M. Sereno, F. De Luca, Phys. Rev. D {\bf 74}, 123009 (2006).
\end{thebibliography}
\end{document}